\documentclass[aps,prl,twocolumn,showcaps,superscriptaddress,showpacs,amsmath,amssymb]{revtex4}

\usepackage{graphicx}
\usepackage{dcolumn}
\usepackage{bm}
\usepackage[dvips]{epsfig}

\newcommand{\unit}[1]{\,\mathrm{#1}}

\begin{document}

\title{Critical Casimir forces in colloidal suspensions on chemically patterned surfaces}

\author{Florian Soyka}
\affiliation{2. Physikalisches Institut, Universit\"at Stuttgart,
Pfaffenwaldring 57, 70550 Stuttgart, Germany}

\author{Olga Zvyagolskaya}
\affiliation{2. Physikalisches Institut, Universit\"at Stuttgart,
Pfaffenwaldring 57, 70550 Stuttgart, Germany}

\author{Christopher Hertlein}
\affiliation{2. Physikalisches Institut, Universit\"at Stuttgart,
Pfaffenwaldring 57, 70550 Stuttgart, Germany}

\author{Laurent Helden}
\affiliation{2. Physikalisches Institut, Universit\"at Stuttgart,
Pfaffenwaldring 57, 70550 Stuttgart, Germany}

\author{Clemens Bechinger}
\affiliation{2. Physikalisches Institut, Universit\"at Stuttgart,
Pfaffenwaldring 57, 70550 Stuttgart, Germany}
\affiliation{Max-Planck-Institut f\"ur Metallforschung,
Heisenbergstr.3, 70569 Stuttgart, Germany}

\begin{abstract}
We investigate the behavior of colloidal particles immersed in a
binary liquid mixture of water and 2,6-lutidine in the presence of a
chemically patterned substrate. Close to the critical point of the
mixture, the particles are subjected to critical Casimir
interactions with force components normal and parallel to the
surface. Because the strength and sign of these interactions can be
tuned by variations in the surface properties and the mixture�s
temperature, critical Casimir forces allow the formation of highly
ordered monolayers but also extend the use of colloids as model
systems.

\end{abstract}

\maketitle

Analogous to the geometrical confinement of
quantum-electrodynamical (QED) vacuum fluctuations between two
parallel metallic plates \cite{casimir48}, the constraint of
concentration fluctuations in fluid mixtures close to their
critical point gives rise to critical Casimir forces acting on the
confining surfaces \cite{fisher1978}. The range of this
interaction is set by the bulk correlation length $\xi$ of the
mixture which diverges when approaching the critical point.
Therefore, critical Casimir forces are sensitive to minute changes
in temperature. Despite several quantitative measurements of such
forces \cite {rafai2007,mukho2000}, it was only recently when the
amplitude of measured critical Casimir forces in quantum- and
classical liquids has been directly compared to theoretical
predictions \cite{garcia1999,ganshin2006,fukuto2005,hertlein2008}.
Direct force measurements of a single colloidal particle above a
flat surface and immersed in a critical water-lutidine mixture
demonstrated, that critical Casimir interactions can easily exceed
multiples of the thermal energy $\unit{k_BT}$ \cite{hertlein2008}.
Accordingly, they offer a versatile opportunity to control the
pair interaction in colloidal suspensions by weak temperature
changes \cite{lu2008, guo2008}. Apart from their exquisite
temperature dependence, critical Casimir forces respond
sensitively to the chemical properties of the confining surfaces.
Depending on whether both surfaces preferentially attract the same
mixture's component or not (symmetric or asymmetric boundary
conditions), attractive or repulsive critical Casimir forces arise
\cite{krech1997, hertlein2008, burk1995}.

So far, experimental investigations of critical Casimir
interactions were limited to homogeneous surfaces (in contrast to
QED Casimir forces \cite{chen2002}) where the corresponding forces
act perpendicular to the confining walls. However, when one or
both surfaces are chemically patterned, also lateral critical
Casimir forces have been predicted \cite{sprenger2006}

In this Letter we experimentally study the interaction between
colloidal particles and chemically patterned substrates immersed
in a binary critical mixture. Close to the critical point lateral
critical Casimir forces lead to the formation of highly ordered
colloidal assemblies whose structure is controlled by the
underlying chemical pattern. This may suggest a novel route for
templated growth of colloidal crystals. At higher particle
concentrations, additional critical Casimir forces between nearby
particle surfaces arise and eventually lead to the formation of
three-dimensional, facetted colloidal islands on the substrate.
Because the particle-substrate and the particle-particle critical
Casimir interactions can be systematically and independently
varied, this also leads to new opportunities in the use of
colloidal model systems to study the growth of islands on surfaces
as this is important for the fabrication of nanostructures
\cite{barth2005}.

We used $2.4\unit{\mu m}$ diameter polystyrene (PS) spheres with a
surface charge of $10\unit{\mu C/cm^2}$ rendering them hydrophilic
which corresponds to ($-$) boundary conditions \cite{galla1992}.
They were suspended in a water-2,6-lutidine (WL) mixture with
critical composition, i.e. a lutidine mass fraction of $c_L^C \cong
0.286$ \cite{beys1985}. WL mixtures have a lower critical demixing
point at $T_C\cong307\unit{K}$ \cite{beys1985}. The Debye screening
length of the mixture was determined with total internal reflection
microscopy (TIRM) to $\kappa^{-1}\cong 12 \unit{nm}$ from the
particle-wall interaction potential between a single colloidal
sphere and a flat glass surface \cite{hertlein2008}. The suspension
was contained in a flat sample cell being assembled from a glass
substrate, a $150\unit{\mu m}$ thick spacer and a glass cover plate
with two filling tubes.

To fabricate glass surfaces with well defined spatial variations
regarding their boundary conditions, they have been first coated
with a monolayer hexamethyldisilazane (HMDS). This was achieved by
exposure to its saturated vapor for several hours in an exicator
\cite{hertlein2008}. This treatment renders the glass surface
hydrophobic with preferential adsorption for lutidine, i.e. ($+$)
boundary condition as confirmed by previous TIRM measurements.
Spatial patterning of the boundary conditions was achieved by
exposure of HMDS coated surfaces to a focussed ion beam (FIB) of
positively charged gallium (Ga) ions. The incident ions locally
remove the HMDS molecules from the surface and thus create
well-defined hydrophilic ($-$) areas. To avoid possible distortion
of the ion beam due to charging effects of the isolating glass
surface, the latter was simultaneously exposed to an electron
beam. With this procedure we created chemical patterns with a
lateral resolution on the order of several tens of nanometers
extending over an area of approx. $400 \mathrm{x} 400 \unit{\mu
m}^2$. Typical parameters for the current, voltage and dose of the
Ga ions were on the order of $1\unit{nA}$, $30\unit{kV}$ and
$26\unit{\mu C/cm^2}$. For the electron beam similar currents but
only voltages on the order of $5\unit{kV}$ were used. After the
substrate patterning, the cell was assembled and the suspension
was inserted. Finally, the filling tubes were sealed with teflon
plugs to avoid evaporation of the mixture. The upper part of the
sample cell was thermally connected to a copper frame (having a
window to allow for optical access) and connected to a flow
thermostat which provided a constant temperature 1 K below $T_C$
with a stability of 0.01K. For precise temperature control, the
cell was actively heated from below with a transparent indium tin
oxide (ITO) coated glass sheet which was operated by a temperature
controller. As thermometers we used two platinum resistors (Pt100)
which were contacted from outside to the sample cell. With this
setup we achieved a temperature stability of about $10\unit{mK}$
over several hours. The entire sample and heating unit was mounted
on the stage of an inverted microscope where particle positions
were determined by digital video microscopy with a spatial
resolution of about $50\unit{nm}$ \cite{baum2005}.

\begin{figure}
  \includegraphics[width=8cm,bb=0 0 2404 1583]{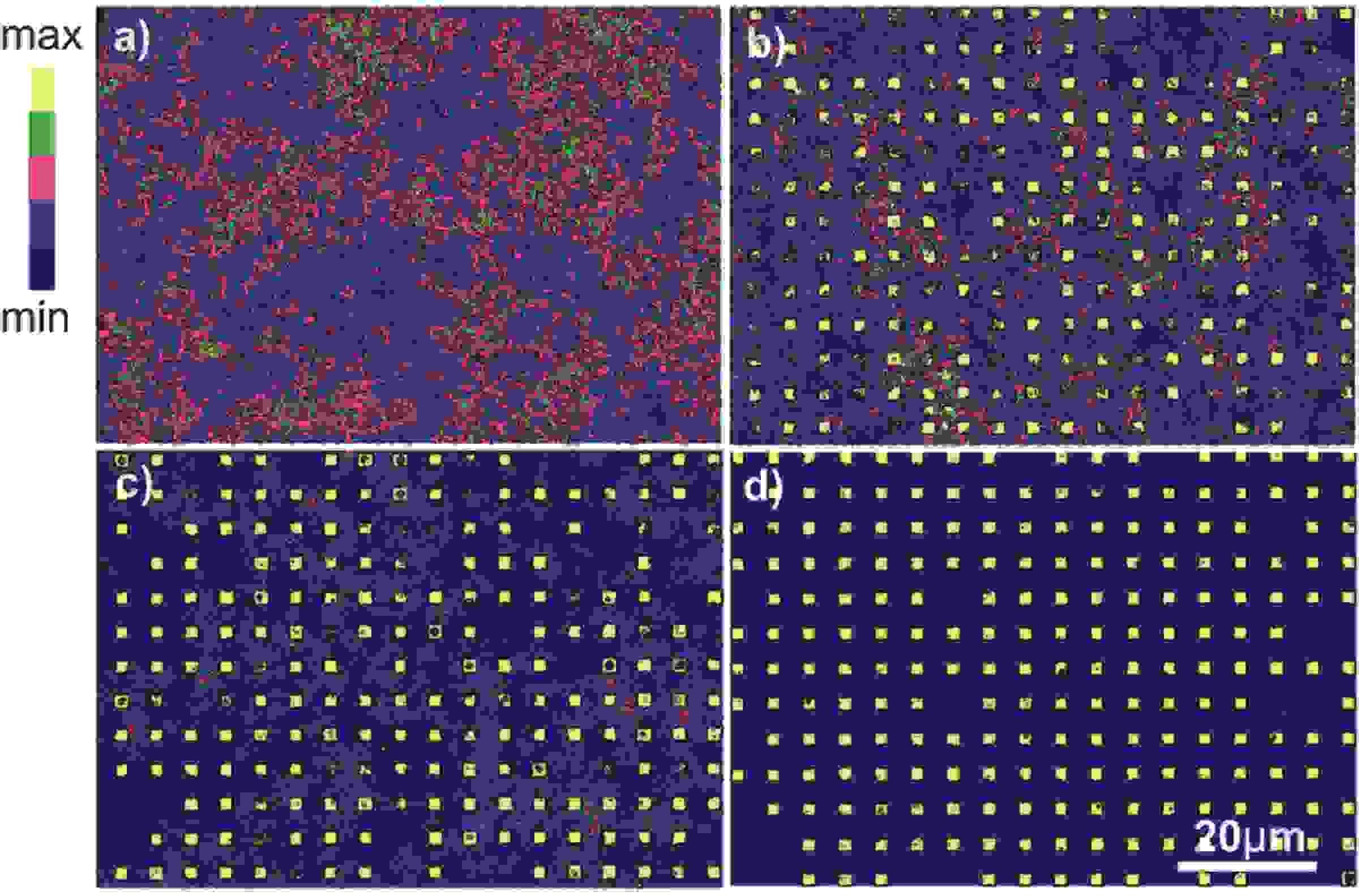}
  \caption{(color online) Averaged particle density distribution $\rho(x,y)$ of a
  diluted colloidal
  suspension of 2.4 $\mu m$ particles in a critical water-lutidine
  mixture in the presence of a chemically patterned substrate.
  $T_c-T=0.72\unit{K}$ (a), $0.25\unit{K}$ (b), $0.23\unit{K}$ (c), and $0.14\unit{K}$ (d).}
  \label{fig_one}
\end{figure}
\begin{figure}
  \includegraphics[width=8cm,bb=0 0 1269 1069]{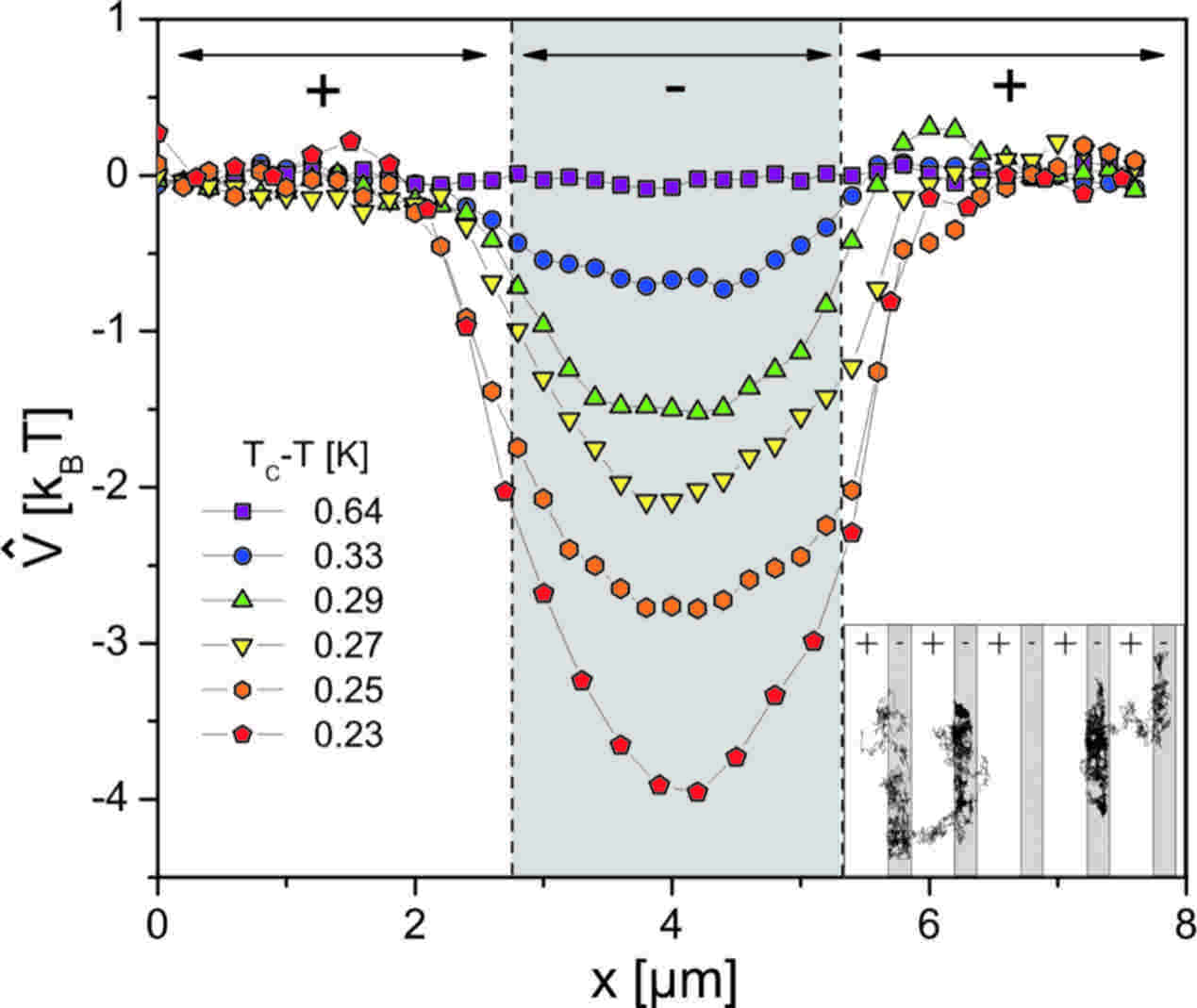}
  \caption{(color online) Temperature-dependent critical Casimir potential acting on a hydrophilic $2.4\unit{\mu m}$ PS particle above
  a chemically striped pattern with alternating  ($-$) and
  ($+$) boundary conditions. Line width are $2.6\unit{\mu m}$ ($-$) and $5.2\unit{\mu m}$ ($+$),
  respectively. Inset: trajectories of two particles recorded over 45 min. on striped chemically patterned
substrate with ($-$) regions indicated in grey.}
  \label{fig_two}
\end{figure}

Fig.~\ref{fig_one} shows the two-dimensional particle distribution
$\rho(x,y)$ of highly diluted colloidal suspension in a WL-mixture
above a chemically patterned substrate. As substrate pattern we
have chosen squares with ($-$) boundary conditions and
$2.6\unit{\mu m}$ side length arranged in a square lattice with
$5.2 \unit{\mu m}$ periodicity. For temperatures
$T_C-T=0.72\unit{K}$ where critical Casimir forces are negligible
\cite{hertlein2008}, the colloidal particles do not respond to the
substrate pattern but are rather evenly distributed across the
field of view (Fig.~\ref{fig_one}a). This demonstrates that
possible lateral variations of the van-der-Waals
\cite{sprenger2006} or electrostatic interactions caused by the
FIB treatment can be neglected under our experimental conditions.
Upon approaching $T_C$, the PS particles become increasingly
localized at the hydrophilic squares while they avoid the
hydrophobic regions as seen in Figs.~\ref{fig_one}b-d. This
behavior is due to gradient forces caused by critical Casimir
interactions which lead to a strongly temperature dependent
attraction of the ($-$) PS particles at the hydrophilic squares
(symmetric boundary conditions) and a repulsion from the
hydrophobic regions ($+$) (asymmetric boundary conditions).
Because the particle size is similar to that of the chemical
substrate pattern, $\rho(x,y)$ eventually resembles the geometry
of the underlying substrate pattern as seen in
Fig.~\ref{fig_one}d.

\begin{figure}
  \includegraphics[width=7cm,bb=0 0 4367 4074]{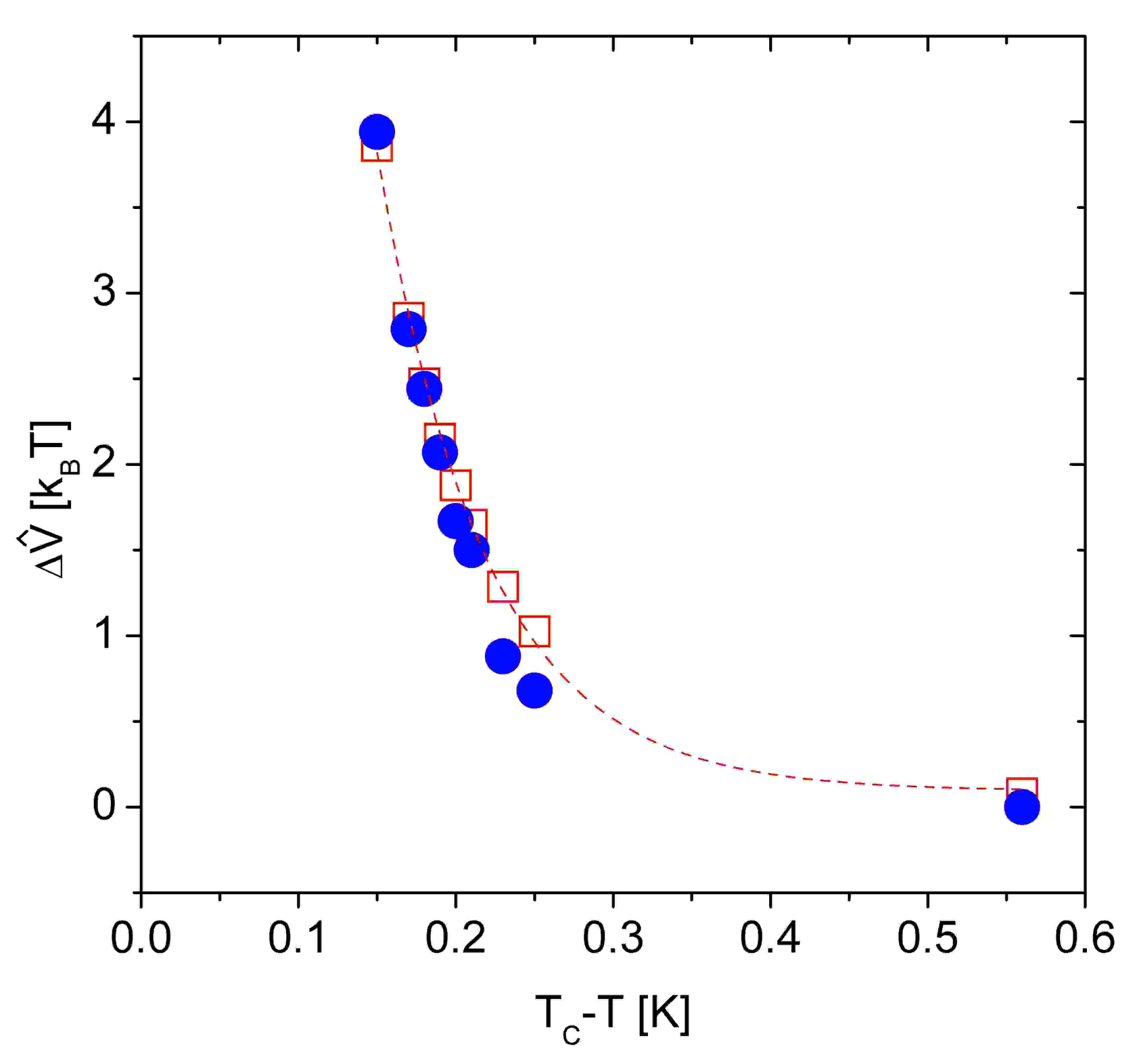}
  \caption{(color online) Temperature dependence of measured (closed symbols) and calculated (open symbols) potential depth of a $2.4\unit{\mu m}$ PS
  particle on a substrate with one-dimensional, periodic alternating boundary conditions.
  The critical temperature has shifted by $80\unit{mK}$ within our experimental accuracy to
  obtain best agreement with the experimental data. The dashed line is a guide to the eye. }
  \label{fig_three}
\end{figure}

To quantify lateral critical Casimir forces, we created substrates
having a one-dimensional periodic chemical pattern, i.e.
hydrophilic ($-$) and hydrophobic ($+$) stripes with $2.6
\unit{\mu m}$ and $5.2 \unit{\mu m}$ width, respectively. Similar
as above, the PS colloids are strongly confined to the ($-$)
stripes when the WL-mixture is heated close to $T_C$ as seen by
the asymmetric shape of the particle trajectories (inset of
Fig.~\ref{fig_two}). The critical Casimir potential $V(x,y,z)$ of
a particle above a patterned substrate depends on its lateral
position $x,y$ and height $z$ above the surface. Since the
$z$-coordinate is not accessible with digital video microcopy, the
measured $\rho(x,y)$ corresponds to the projection of the
three-dimensional particle distribution onto the $xy$-plane. For
the one-dimensional surface pattern as considered here, these data
can be further projected onto the $x$-axis (being perpendicular to
the direction of the stripes). From the resulting one-dimensional
particle distribution $\hat{\rho}(x)$ we define the effective
one-dimensional critical Casimir potential via the Boltzmann
factor
\begin{equation}\label{eq_minusone}
     \frac{\hat{V}(x)}{\unit{k_BT}}=-\ln{\hat{\rho}(x)}+c
 \end{equation}
with $c$ an arbitrary constant. In Fig.~\ref{fig_two} we show the
results for a single period as a function of temperature. It is
clearly seen that upon approaching $T_C$, the critical Casimir
potential becomes increasingly attractive at the stripes with
($-$) boundary condition. The measured nonlinear temperature
dependence of the potential depth $\Delta\hat{V}$ is shown in
Fig.~\ref{fig_three} as closed symbols. We have limited these
measurements to temperatures where $\Delta\hat{V}< 4\unit{k_BT}$
to allow for occasional particle escapes from the potential wells
during the measuring time which are required to sample the entire
energy landscape. It should be mentioned, however, that much
deeper potentials can be obtained by further approaching $T_C$.

For a quantitative estimate, we assumed that $\Delta\hat{V}$ is
given by
\begin{equation}\label{eq_zero}
    \Delta\hat{V}\cong V_{(- -)}-V_{(+ -)}
 \end{equation}
where $V_{(- -)}$ and $V_{(+ -)}$ are the critical Casimir
potentials of the colloid above infinite homogeneous substrates
with ($-$) and ($+$) boundary condition, respectively. These
potentials only depend on the corresponding particle heights
$z_{(- -)}$ and $z_{(+ -)}$. Although Eq. (\ref{eq_zero}) ignores
the finite width of the stripes (which is only slightly larger
than the particle diameter), this simplification is justified here
as can be seen by the almost vanishing curvature of $\hat{V}(x)$
near its minimum. The critical Casimir potential of a colloidal
sphere with radius $R$ at height $z$ above a homogeneous surface
is given by \cite{hertlein2008}

 \begin{equation}\label{eq_one}
    \frac{V}{\unit{k_BT}}=\frac{R}{z}\vartheta\left( \frac{z}{\xi}\right)
 \end{equation}

with $\vartheta$ the corresponding universal scaling function for
symmetric and asymmetric boundary condition which have been
inferred from recent Monte Carlo simulations for classical binary
mixtures \cite{vasilyev2007,hertlein2008}. The correlation length
is given by $\xi = \xi_0(\frac{T_C-T}{T_C})^{-0.63}$  with
$\xi_0\approx 0.2\unit{nm}$ for water-lutidine mixtures as
determined experimentally \cite{guelari1972,hertlein2008}. What
remains for the calculation of $\Delta\hat{V}$ from
Eqs.(\ref{eq_zero}) and (\ref{eq_one}) are the
temperature-dependent particle heights of a 2.4$\unit{\mu m}$ PS
particle on a substrate with ($- -$) and ($+ -$) boundary
condition. The height distributions have been measured with TIRM
with an accuracy of $\pm 30\unit{nm}$ and yield
temperature-dependent mean values of $\langle z\rangle_{(-
-)}=0.12\unit{\mu m}$ and $0.11 \unit{\mu m}\leq\langle
z\rangle_{(+ -)}\leq 0.2 \unit{\mu m}$. Note, that within our
experimental resolution, $\langle z\rangle_{(- -)}$ does not vary
with temperature due to the strong electrostatic repulsion at
small distances. The calculated temperature-dependence of
$\Delta\hat{V}$ is shown as open symbols in Fig. \ref{fig_three}.
Best agreement with the experimental data is obtained when
assuming that ${T_C}$ is $80\unit{mK}$ lower than the temperature
where the onset of phase separation (being identified as ${T_C}$)
has been experimentally observed. This shift is within our error
in determining the critical temperature with this procedure. The
good agreement in Fig. \ref{fig_three} also confirms that the
above assumption (Eq. (\ref{eq_zero})) is indeed reasonable under
our conditions.

\begin{figure}
  \includegraphics[width=8cm,bb=0 0 1000 1051]{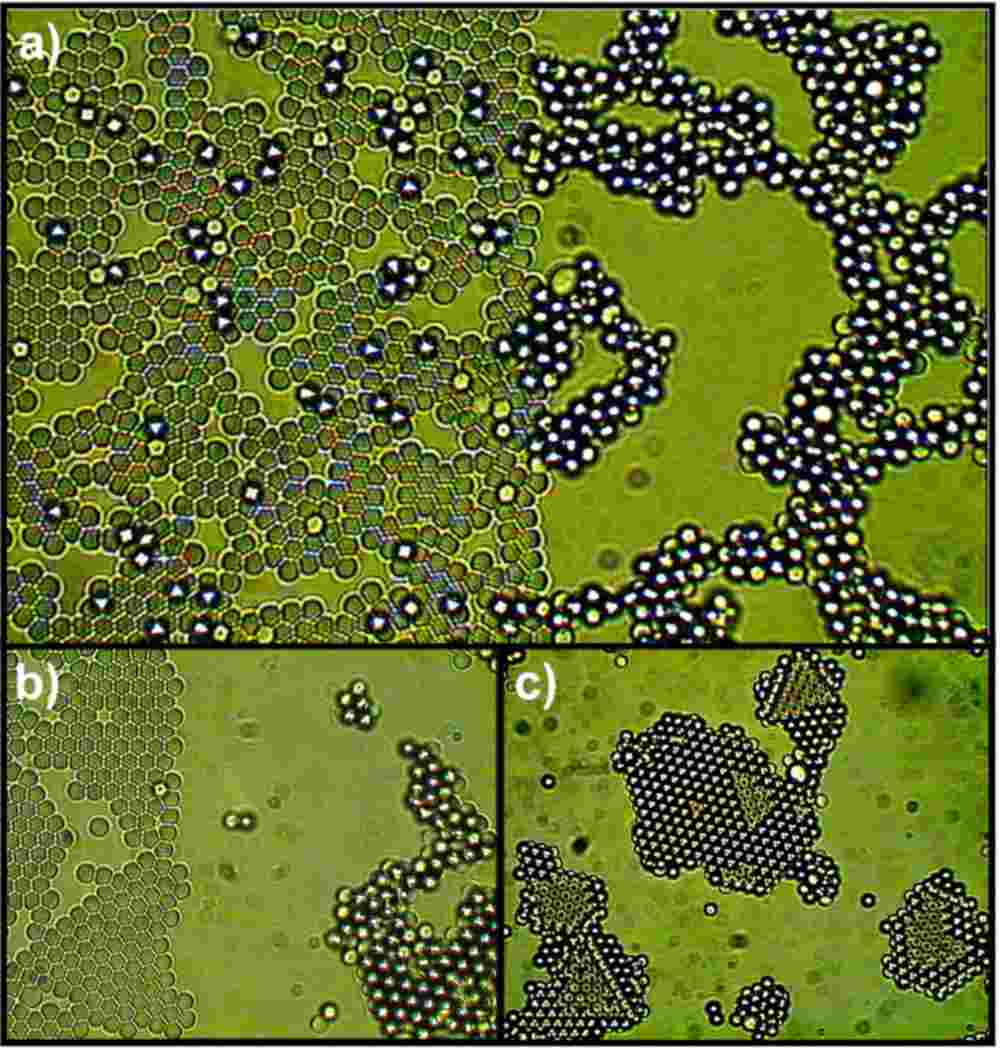}
  \caption{(color online) (a) Snapshot of a colloidal suspension above a substrate with a chemical
  step, i.e.
  ($-$) and ($+$) boundary conditions at the
  left and right part of the sample. Due to attractive (left) and
  repulsive (right)
  critical Casimir forces between the particles and the substrate,
  a monolayer forms on the left while at the right side the particles arrange in three-dimensional clusters.
(b) Detail of the region at the chemical
  interface after the system was kept for several hours at constant temperature. (c) Optical image of the faceted islands
  which form on the hydrophobic substrate when the sample is kept for several hours at constant temperature.}
  \label{fig_four}
\end{figure}

When further increasing the particle density, critical
concentration fluctuations become also confined between adjacent
colloidal surfaces. This leads - in addition to the forces between
the particles and the substrate - to critical Casimir interactions
between adjacent particles. Within the Derjaguin approximation its
amplitude is one-half of Eq. (\ref{eq_one}) \cite{schles2003}. To
study the interplay of those forces we created a substrate with a
single chemical step, i.e. with ($-$) boundary conditions on the
left and ($+$) on the right side. As can be seen in
Fig.~\ref{fig_four} the colloidal particles in a critical
water-lutidine mixture form rather different structures on each
side of the substrate. On the left side, all boundary conditions
(particle-particle and particle-substrate) are symmetric ($- -$)
thus leading to attractive interactions. Upon approaching the
critical temperature, this leads essentially to a loosely-packed
single layer of particles as seen on the left side of
Fig.~\ref{fig_four}a. Because of vertical particle fluctuations,
some colloids (appearing as bright spots) are initially located on
top of this layer when increasing the temperature towards ${T_C}$.
However, when keeping the system at constant temperature, most of
these particles laterally diffuse until they reach a void where
they merge into the first layer (Fig.~\ref{fig_four}b, left).


The situation on the hydrophobic part of the sample (right side of
Fig.~\ref{fig_four}a) is different, because here the particles are
repelled from the substrate (($+ -$) boundary conditions). When
approaching the critical temperature, the particles initially form
a three-dimensional rather open structure. At larger timescales
and constant temperature, however, the clusters rearrange and
eventually form faceted close-packed islands with heights up to
three particle diameters (Fig.~\ref{fig_four}c). At very large
timescales (several hours) even the aggregation of entire clusters
has been observed. We expect that for other parameters where the
critical Casimir interaction energy largely exceeds $\unit{k_BT}$,
surface diffusion of particles becomes reduced. Accordingly, this
should then lead to fractal or dendritic structures.

To conclude, we have reported on the use of critical Casimir
forces as a novel and robust approach for colloidal assembly on
chemically patterned surfaces. Since these forces can be
controlled by variations in the surface properties and the
mixtures temperature, critical Casimir forces lead to a rich phase
behavior. In contrast to other aggregation mechanisms, critical
Casimir forces are fully reversible which allows to thermally
anneal the obtained structures. This will result in a largely
reduced defect density being important for technical applications
\cite{velev1999}\cite{albrecht2005}. Since critical Casimir forces
are not restricted to micron-sized particles, the principle as
discussed here should be also applicable on smaller length scales.
It should be also mentioned that chemical patterns can be obtained
by other techniques than a FIB, e.g. by self-assembled monolayers
deposited on gold-coated substrates with microcontact-printing
\cite{kumar1993}.

\begin{acknowledgments}
We thank U. Eigenthaler, A. Breitling and M. Hirscher for
technical assistance during creation of chemical substrate
patterning with the FIB, U. Nellen for TIRM measurements and A.
Gambassi, L. Harnau and S. Dietrich for helpful discussions.
\end{acknowledgments}

\bibliographystyle{unsrt}

\begin{thebibliography} {2}
\expandafter\ifx\csname natexlab\endcsname\relax\def\natexlab#1{#1}\fi \expandafter\ifx\csname bibnamefont\endcsname\relax
  \def\bibnamefont#1{#1}\fi
\expandafter\ifx\csname bibfnamefont\endcsname\relax
  \def\bibfnamefont#1{#1}\fi
\expandafter\ifx\csname citenamefont\endcsname\relax
  \def\citenamefont#1{#1}\fi
\expandafter\ifx\csname url\endcsname\relax
  \def\url#1{\texttt{#1}}\fi
\expandafter\ifx\csname urlprefix\endcsname\relax\def\urlprefix{URL }\fi \providecommand{\bibinfo}[2]{#2}
\providecommand{\eprint}[2][]{\url{#2}}

 \bibitem[{\citenamefont{Casimir}(1948)}]{casimir48}
\bibinfo{author}{\bibfnamefont{H. B. G.} \bibnamefont{Casimir}},
\bibinfo{journal}{Proc. Kon. Nederl. Akad. Wet} \textbf{\bibinfo{volume}{B51}},
  \bibinfo{pages}{793} (\bibinfo{year}{1948}).

  \bibitem[{\citenamefont{Fisher}(1978)}]{fisher1978}
\bibinfo{author}{\bibfnamefont{M. E.} \bibnamefont{Fischer}},
\bibnamefont{and}
\bibinfo{author}{\bibfnamefont{P. G.} \bibnamefont{de Gennes}},
\bibinfo{journal}{C. R. Acad. Sci. paris} \textbf{\bibinfo{volume}{B287}},
  \bibinfo{pages}{207} (\bibinfo{year}{1978}).


\bibitem[{\citenamefont{Mukhopadhyay}(2000)}]{mukho2000}
\bibinfo{author}{\bibfnamefont{A.} \bibnamefont{Mukhopadhyay}},
\bibnamefont{and}
\bibinfo{author}{\bibfnamefont{B. M.} \bibnamefont{Law}},
\bibinfo{journal}{Phys. Rev. Lett.} \textbf{\bibinfo{volume}{83}},
  \bibinfo{pages}{772} (\bibinfo{year}{1999}).


   \bibitem[{\citenamefont{Rafa\"{\i}}(2007)}]{rafai2007}
\bibinfo{author}{\bibfnamefont{S.} \bibnamefont{Rafai}},
\bibinfo{author}{\bibfnamefont{D.} \bibnamefont{Bonn}},
\bibnamefont{and}
\bibinfo{author}{\bibfnamefont{J.} \bibnamefont{Meunier}},
\bibinfo{journal}{Physica A} \textbf{\bibinfo{volume}{386}},
  \bibinfo{pages}{31} (\bibinfo{year}{2007}).

 \bibitem[{\citenamefont{Garcia}(1999)}]{garcia1999}
\bibinfo{author}{\bibfnamefont{R.} \bibnamefont{Garcia}},
\bibnamefont{and}
\bibinfo{author}{\bibfnamefont{M. H. W.} \bibnamefont{Chan}},
\bibinfo{journal}{Phys. Rev. Lett.} \textbf{\bibinfo{volume}{83}},
  \bibinfo{pages}{1187} (\bibinfo{year}{1999}).

   \bibitem[{\citenamefont{Ganshin}(2006)}]{ganshin2006}
\bibinfo{author}{\bibfnamefont{A.} \bibnamefont{Ganshin}},
\bibinfo{author}{\bibfnamefont{S.} \bibnamefont{Scheidemantel}},
\bibinfo{author}{\bibfnamefont{R.} \bibnamefont{Garcia}},
\bibnamefont{and}
\bibinfo{author}{\bibfnamefont{M. H. W.} \bibnamefont{Chan}},
\bibinfo{journal}{Phys. Rev. Lett. } \textbf{\bibinfo{volume}{97}},
  \bibinfo{pages}{075301} (\bibinfo{year}{2006}).

\bibitem[{\citenamefont{Fukuto}(2005)}]{fukuto2005}
\bibinfo{author}{\bibfnamefont{M.} \bibnamefont{Fukuto}},
\bibinfo{author}{\bibfnamefont{Y. F.} \bibnamefont{Yano}},
\bibnamefont{and}
\bibinfo{author}{\bibfnamefont{P. S.} \bibnamefont{Pershan}},
\bibinfo{journal}{Phys. Rev. Lett.} \textbf{\bibinfo{volume}{94}},
  \bibinfo{pages}{135702} (\bibinfo{year}{2005}).

\bibitem[{\citenamefont{Hertlein}(2008)}]{hertlein2008}
\bibinfo{author}{\bibfnamefont{C.} \bibnamefont{Hertlein}},
\bibinfo{author}{\bibfnamefont{L.} \bibnamefont{Helden}},
\bibinfo{author}{\bibfnamefont{A.} \bibnamefont{Gambassi}},
\bibinfo{author}{\bibfnamefont{S.} \bibnamefont{Dietrich}},
\bibnamefont{and}
\bibinfo{author}{\bibfnamefont{C.} \bibnamefont{Bechinger}},
\bibinfo{journal}{Nature} \textbf{\bibinfo{volume}{451}},
  \bibinfo{pages}{172} (\bibinfo{year}{2008}).

 \bibitem[{\citenamefont{Guo}(20008)}]{guo2008}
\bibinfo{author}{\bibfnamefont{H.} \bibnamefont{Guo}},
\bibinfo{author}{\bibfnamefont{T.} \bibnamefont{Narayanan}},
\bibinfo{author}{\bibfnamefont{M.} \bibnamefont{Sztuchi}},
\bibinfo{author}{\bibfnamefont{P.} \bibnamefont{Schall}}, \bibnamefont{and}
\bibinfo{author}{\bibfnamefont{G. H.} \bibnamefont{Wegdam}},
\bibinfo{journal}{Phys. Rev. Lett.} \textbf{\bibinfo{volume}{100}},
  \bibinfo{pages}{188303} (\bibinfo{year}{2008}).

\bibitem[{\citenamefont{Lu}(2008)}]{lu2008}
\bibinfo{author}{\bibfnamefont{X. H.} \bibnamefont{Lu}},
\bibinfo{author}{\bibfnamefont{S. G. J.} \bibnamefont{Mochrie}},
\bibinfo{author}{\bibfnamefont{S.} \bibnamefont{Narayanan}},
\bibinfo{author}{\bibfnamefont{A. R.} \bibnamefont{Sandy}},
\bibnamefont{and}
\bibinfo{author}{\bibfnamefont{M.} \bibnamefont{Sprung}},
\bibinfo{journal}{Phys. Rev. Lett.} \textbf{\bibinfo{volume}{100}},
  \bibinfo{pages}{045701} (\bibinfo{year}{2008}).

\bibitem[{\citenamefont{Krech}(1997)}]{krech1997}
\bibinfo{author}{\bibfnamefont{M.} \bibnamefont{Krech}},
\bibinfo{journal}{Phys. Rev. E} \textbf{\bibinfo{volume}{56}},
  \bibinfo{pages}{1642} (\bibinfo{year}{1997}).

 \bibitem[{\citenamefont{Burkhardt}(1995)}]{burk1995}
\bibinfo{author}{\bibfnamefont{T. W.} \bibnamefont{Burkhardt}},
\bibnamefont{and}
\bibinfo{author}{\bibfnamefont{E.} \bibnamefont{Eisenriegler}},
\bibinfo{journal}{Phys. Rev. Lett.} \textbf{\bibinfo{volume}{74}},
  \bibinfo{pages}{3189} (\bibinfo{year}{1995}).

 \bibitem[{\citenamefont{Chen}(2002)}]{chen2002}
\bibinfo{author}{\bibfnamefont{F.} \bibnamefont{Chen}},
\bibinfo{author}{\bibfnamefont{U.} \bibnamefont{Mohideen}},
\bibinfo{author}{\bibfnamefont{G.L.} \bibnamefont{Klimchitskaya}},
\bibnamefont{and}
\bibinfo{author}{\bibfnamefont{V.M.} \bibnamefont{Mostepanenko}},
\bibinfo{journal}{Phys. Rev. Lett.} \textbf{\bibinfo{volume}{88}},
  \bibinfo{pages}{101801} (\bibinfo{year}{2002}).

\bibitem[{\citenamefont{Sprenger}(2006)}]{sprenger2006}
\bibinfo{author}{\bibfnamefont{M.} \bibnamefont{Sprenger}},
\bibinfo{author}{\bibfnamefont{F.} \bibnamefont{Schlesener}},
\bibnamefont{and}
\bibinfo{author}{\bibfnamefont{S.} \bibnamefont{Dietrich}},
\bibinfo{journal}{J. Chem. Phys.} \textbf{\bibinfo{volume}{124}},
  \bibinfo{pages}{134703} (\bibinfo{year}{2006}).

\bibitem[{\citenamefont{Barth}(1993)}]{barth2005}
\bibinfo{author}{\bibfnamefont{J. V.} \bibnamefont{Barth}},
\bibinfo{author}{\bibfnamefont{G.} \bibnamefont{Costantini}},
\bibnamefont{and}
\bibinfo{author}{\bibfnamefont{K.} \bibnamefont{Kern}},
\bibinfo{journal}{Nature} \textbf{\bibinfo{volume}{437}},
  \bibinfo{pages}{671} (\bibinfo{year}{2005}).

\bibitem[{\citenamefont{Gallagher}(1992)}]{galla1992}
\bibinfo{author}{\bibfnamefont{P. D.} \bibnamefont{Gallagher}},
\bibinfo{author}{\bibfnamefont{M. L.} \bibnamefont{Kurnaz}},
\bibnamefont{and}
\bibinfo{author}{\bibfnamefont{ J. V. } \bibnamefont{Maher}},
\bibinfo{journal}{Phys. Rev. A} \textbf{\bibinfo{volume}{46}},
  \bibinfo{pages}{7750} (\bibinfo{year}{1992}).


 \bibitem[{\citenamefont{Beysens}(1985)}]{beys1985}
\bibinfo{author}{\bibfnamefont{D.} \bibnamefont{Beysens}},
\bibnamefont{and}
\bibinfo{author}{\bibfnamefont{D.} \bibnamefont{Esteve}},
\bibinfo{journal}{Phys. Rev. Lett.} \textbf{\bibinfo{volume}{54}},
  \bibinfo{pages}{2123} (\bibinfo{year}{1985}).


 \bibitem[{\citenamefont{Baumgartl}(2005)}]{baum2005}
\bibinfo{author}{\bibfnamefont{J.} \bibnamefont{Baumgartl}},
\bibnamefont{and}
\bibinfo{author}{\bibfnamefont{C.} \bibnamefont{Bechinger}},
\bibinfo{journal}{Europhys. Lett.} \textbf{\bibinfo{volume}{71}},
  \bibinfo{pages}{487} (\bibinfo{year}{2005}).

\bibitem[{\citenamefont{Vasilyev}(2007)}]{vasilyev2007}
\bibinfo{author}{\bibfnamefont{O.} \bibnamefont{Vasilyev}},
\bibinfo{author}{\bibfnamefont{A.} \bibnamefont{Gambassi}},
\bibinfo{author}{\bibfnamefont{A.} \bibnamefont{Maciolek}},
\bibnamefont{and}
\bibinfo{author}{\bibfnamefont{S.} \bibnamefont{Dietrich}}
\bibinfo{journal}{Europhys. Lett.} \textbf{\bibinfo{volume}{80}},
  \bibinfo{pages}{60009} (\bibinfo{year}{2007}).

\bibitem[{\citenamefont{G�lari}(1972)}]{guelari1972}
\bibinfo{author}{\bibfnamefont{E.} \bibnamefont{G\"{u}lari}},
\bibinfo{author}{\bibfnamefont{A. F.} \bibnamefont{Collings}},
\bibnamefont{and}
\bibinfo{author}{\bibfnamefont{R. L.} \bibnamefont{Schmidt}},
\bibinfo{journal}{J. Chem. Phys.} \textbf{\bibinfo{volume}{56}},
  \bibinfo{pages}{6169} (\bibinfo{year}{1972}).

\bibitem[{\citenamefont{Schlesener}(2003)}]{schles2003}
\bibinfo{author}{\bibfnamefont{F.} \bibnamefont{Schlesener}},
\bibinfo{author}{\bibfnamefont{A.} \bibnamefont{Hanke}},
\bibnamefont{and}
\bibinfo{author}{\bibfnamefont{S.} \bibnamefont{Dietrich}},
\bibinfo{journal}{J. Stat. Phys.} \textbf{\bibinfo{volume}{110}},
  \bibinfo{pages}{981} (\bibinfo{year}{2003}).




 \bibitem[{\citenamefont{Velev}(1999)}]{velev1999}
\bibinfo{author}{\bibfnamefont{O. D.} \bibnamefont{Velev}},
\bibnamefont{and}
\bibinfo{author}{\bibfnamefont{E. W.} \bibnamefont{Kaler}},
\bibinfo{journal}{Langmuir} \textbf{\bibinfo{volume}{15}},
  \bibinfo{pages}{3693} (\bibinfo{year}{1999}).

\bibitem[{\citenamefont{Albrecht}(2005)}]{albrecht2005}
\bibinfo{author}{\bibfnamefont{A.} \bibnamefont{Albrecht}},
\bibinfo{author}{\bibfnamefont{G.} \bibnamefont{Hu}},
\bibinfo{author}{\bibfnamefont{I. L.} \bibnamefont{Guhr}},
\bibinfo{author}{\bibfnamefont{T. C.} \bibnamefont{Ulbrich}},
\bibinfo{author}{\bibfnamefont{J.} \bibnamefont{Boneberg}},
\bibinfo{author}{\bibfnamefont{P.} \bibnamefont{Leiderer}},
\bibnamefont{and}
\bibinfo{author}{\bibfnamefont{G.} \bibnamefont{Schatz}},
\bibinfo{journal}{Nature Materials } \textbf{\bibinfo{volume}{4}},
  \bibinfo{pages}{203} (\bibinfo{year}{2005}).



 \bibitem[{\citenamefont{Kumar }(1993)}]{kumar1993}
\bibinfo{author}{\bibfnamefont{A.} \bibnamefont{Kumar}},
\bibnamefont{and}
\bibinfo{author}{\bibfnamefont{G. M.} \bibnamefont{Whitesides}},
\bibinfo{journal}{Appl. Phys. Lett.} \textbf{\bibinfo{volume}{63}},
  \bibinfo{pages}{2002} (\bibinfo{year}{1993}).














\end{thebibliography}

\end{document}